%
%
%

\def\oddpage{{\tt preliminary draft \hfil \jobname \hfil \today}}
\def\evenpage{{\tt \today \hfil\jobname \hfil preliminary draft}}
\def\titlepage{{\tt preliminary draft \hfill \jobname}}
\def\draft{\baselineskip = 16pt plus 2pt minus 1pt
   \hsize = 17.0 truecm \vsize = 24.7 truecm
   \hoffset=-4 truemm   
    \overfullrule = 5pt
    \headline{\ifnum\count0>1\ifodd\count0\oddpage\else%
    \evenpage\fi\else\titlepage\fi}
}
\overfullrule=0pt             
\hoffset=-3 true mm           
\voffset=9.5 true mm

%
%
%
%

\newif\iflabels

\newif\ifreference


\newcount\secnum    \global\secnum=0
\newcount\subsecnum \global\subsecnum=0
\newcount\eqnum     \global\eqnum=0
\newcount\citenum   \global\citenum=0
\newcount\fignum    \global\fignum=0
\newcount\tablenum  \global\tablenum=0
\newcount\remarknum    \global\remarknum=0

\newif\ifndouble
\def\doublenumbers{\ndoubletrue\gdef\therunningsection{\the\secnum}}

\def\ifundefined#1{\expandafter\ifx\csname#1\endcsname\relax}
\def\strip#1>{}


\def\cref#1{\ifundefined{@c@#1}\immediate\write16{ --> \string\cref{#1}
    not defined !!!}
    \expandafter\xdef\csname@c@#1\endcsname{??}\fi\csname@c@#1\endcsname}

\def\eqref#1{\ifundefined{@eq@#1}\immediate\write16{ --> \string\eqref{#1}
    not defined !!!}
    \expandafter\xdef\csname@eq@#1\endcsname{??}\fi\csname@eq@#1\endcsname}

\def\sref#1{\ifundefined{@s@#1}\immediate\write16{ --> \string\sref{#1}
    not defined !!!}
    \expandafter\xdef\csname@s@#1\endcsname{??}\fi\csname@s@#1\endcsname}

\def\figref#1{\ifundefined{@f@#1}\immediate\write16{ -->
    \string\figref{#1} not defined !!!}
    \expandafter\xdef\csname@f@#1\endcsname{??}\fi\csname@f@#1\endcsname}

\def\tableref#1{\ifundefined{@f@#1}\immediate\write16{ -->
    \string\tableref{#1} not defined !!!}
    \expandafter\xdef\csname@f@#1\endcsname{??}\fi\csname@f@#1\endcsname}

\newdimen\beforesecskip  \beforesecskip=\baselineskip
\newdimen\aftersecskip   \aftersecskip=0pt

\font\sectionfont = cmbx10 at 11 pt

\def\Section#1#2{            
    \global\advance\secnum by 1\ifndouble\global\eqnum=0\fi
    \global\subsecnum=0
    \xdef\therunningsection{\the\secnum}
    \def\usenthrow{1}\ifundefined{@s@#1}\def\usenthrow{2}\fi
    \expandafter\ifx\csname@s@#1\endcsname\therunningsection\def\usenthrow{2}\fi
    \ifodd\usenthrow\immediate\write16
      { --> Possible reference error in \string\sref{#1} }\fi
    \expandafter\xdef\csname@s@#1\endcsname{\therunningsection}
    \immediate\write16{\therunningsection. #2}
    \goodbreak\vskip\beforesecskip\noindent%
    \iflabels
          \llap{\tt #1\quad}%
    \fi
    {\sectionfont\the\secnum.\enspace#2}\par\nobreak\noindent\ignorespaces}

\def\subsection#1{\global\advance\subsecnum by 1%
    \xdef\therunningsubsection{\the\subsecnum}%
    \medbreak\smallskip
    \noindent{\sl\the\secnum.\the\subsecnum.\enspace #1}%
    \nobreak\noindent}

\def\section#1#2{\Section{#2}{#1}\par\noindent\ignorespaces}

\def\asection#1{\immediate\write16{#1}%
    \goodbreak\vskip\beforesecskip
    \noindent{\sectionfont#1}\par\nobreak\noindent\ignorespaces}


\def\eqlabel#1{\global\advance\eqnum by 1
    \ifndouble\xdef\anumber{\therunningsection.\the\eqnum}
       \else\xdef\anumber{\the\eqnum}\fi
    \def\usenthrow{1}\ifundefined{@eq@#1}\def\usenthrow{2}\fi
    \expandafter\ifx\csname@eq@#1\endcsname\anumber\def\usenthrow{2}\fi
    \ifodd\usenthrow\immediate\write16
       { --> Possible reference error in \string\eqref{#1} }\fi
    \expandafter\xdef\csname@eq@#1\endcsname{\anumber}
    \ifndouble
       \def\usenthrow{\expandafter\strip\meaning\therunningsection.\the\eqnum}
       \else\def\usenthrow{\the\eqnum}\fi
}

\def\autoeqno#1{\eqlabel{#1}\eqno(\csname@eq@#1\endcsname)
    \iflabels \rlap{\quad\tt #1} \fi
}
\def\autoleqno#1{\eqlabel{#1}\leqno(\csname@eq@#1\endcsname)
    \iflabels \llap{\tt #1 \qquad} \fi
}

\def\therefs{}
\def\bibitem#1#2\par{
    \global\advance\citenum by 1
    \xdef\citation{\the\citenum}
    \def\usenthrow{1}\ifundefined{@c@#1}\def\usenthrow{2}\fi
    \expandafter\ifx\csname@c@#1\endcsname\citation\def\usenthrow{2}\fi
    \ifodd\usenthrow\immediate\write16
      { --> Possible reference error in \string\cref{#1} }\fi
    \expandafter\xdef\csname@c@#1\endcsname{\citation}
\iflabels
     \ifnum\citenum = 1\global\xdef\therefs{\par\noindent\llap{\tt#1\qquad}%
          \ignorespaces#2\par}
     \else 
          \global\xdef\oldrefs{\therefs}
          \global\xdef\therefs{\oldrefs\par\noindent\llap{\tt#1\qquad}%
          \ignorespaces#2\par}
     \fi
\else
     \ifnum\citenum = 1\global\xdef\therefs{\item{[\citation]} #2\par }
     \else 
          \global\xdef\oldrefs{\therefs}
          \global\xdef\therefs{\oldrefs\item{[\citation]} #2\par }%
     \fi
\fi
}


\def\cite#1{\hbox{[\cref{#1}]}}

\newcount\refcount
\refcount=1
\def\listrefs{\frenchspacing
    \asection{References}\par
    \iflabels
          {
          \everypar{\hang\textindent{[\the\refcount]}
          \global\advance\refcount by 1\relax}\therefs
          }
    \else
          \therefs
    \fi
    \nonfrenchspacing}



\newdimen\captionwidth
\captionwidth = \hsize
\advance\captionwidth by -2\parindent
\newbox\captionbox

\def\figure#1#2#3{
    \global\advance\fignum by 1
    \xdef\afigure{\the\fignum}
    \def\usenthrow{1}\ifundefined{@f@#1}\def\usenthrow{2}\fi
    \expandafter\ifx\csname@f@#1\endcsname\afigure\def\usenthrow{2}\fi
    \ifodd\usenthrow\immediate\write16
      { --> Possible reference error in \string\figref{#1} }\fi
    \expandafter\xdef\csname@f@#1\endcsname{\afigure}
    \ifnum\fignum = 1\global\xdef\thefigs{\item{Fig.\ \afigure.} #2\ }
    \else%
    \global\xdef\oldfigs{\thefigs}%
    \global\xdef\thefigs{\oldfigs\item{Fig.\ \afigure:} #2\ }%
    \fi%
     \goodbreak\midinsert
     \ifx\epsfbox\undefined
          \immediate\write16{ Fig. \afigure: ignored }
          \noindent\hrule\par\vskip 1cm \noindent\hrule
     \else\immediate\write16{ Fig. \afigure.}
          \center{#3}          
     \fi
     \smallskip
     \setbox\captionbox=\hbox{Figure \afigure: \ignorespaces#2}
     \iflabels
          \ifdim \wd\captionbox < \captionwidth
               \noindent\llap{\tt#1\quad}\centerline{Figure \afigure: \ignorespaces#2}
          \else
               \noindent
               \llap{\tt#1\quad}{\narrower\noindent Figure \afigure: \ignorespaces#2\par}
          \fi
     \else
          \ifdim \wd\captionbox < \captionwidth \centerline{Figure \afigure: \ignorespaces#2}
          \else {\narrower\noindent Figure \afigure: \ignorespaces#2\par}
          \fi
     \fi
     \endinsert
}


\def\table#1#2#3{
    \global\advance\tablenum by 1
    \xdef\atable{\the\tablenum}
    \def\usenthrow{1}\ifundefined{@f@#1}\def\usenthrow{2}\fi
    \expandafter\ifx\csname@f@#1\endcsname\atable\def\usenthrow{2}\fi
    \ifodd\usenthrow\immediate\write16
      { --> Possible reference error in \string\tableref{#1} }\fi
    \expandafter\xdef\csname@f@#1\endcsname{\atable}
    \ifnum\tablenum = 1\global\xdef\thetables{\item{Table \atable.} #2\ }
    \else%
    \global\xdef\oldtables{\thetables}%
    \global\xdef\thetables{\oldtables\item{Table \atable.} #2\ }%
    \fi%
     \goodbreak\midinsert
     \immediate\write16{ Table \atable.}
     \setbox\captionbox=\hbox{Table \atable. \ignorespaces#2}
     \iflabels
          \ifdim \wd\captionbox < \captionwidth
               \noindent\llap{\tt#1\quad}\centerline{Table \atable. \ignorespaces#2}
          \else
               \noindent
               \llap{\tt#1\quad}{\narrower\noindent Table \atable. \ignorespaces#2\par}
          \fi
     \else
          \ifdim \wd\captionbox < \captionwidth \centerline{Table \atable. \ignorespaces#2}
          \else {\narrower\noindent Table \atable. \ignorespaces#2\par}
          \fi
     \fi
     \medskip
     \let\\=\cr
     \centerline{\vbox{\offinterlineskip\halign{\strut\ignorespaces#3}}}
     \medskip
     \endinsert
}


\def\hline{\noalign{\hrule}}
\def\today{\ifcase\month\or January\or February\or
   March\or April\or May\or June\or July\or August\or September\or
   October\or November\or December\fi
   \space\number\day, \number\year}
\def\frac#1#2{{#1\over#2}}

\def\text#1{{\rm #1}}
\def\degs{\ifmmode {}^\circ \else ${}^\circ$ \fi} 
\def\[{\begingroup$$\let\\=\cr}
\def\]{$$\endgroup\ignorespaces}
\def\\{\hfil\break}
\def\){\hfill\break}

\def\roughly#1{\mathrel{\raise.3ex\hbox{$#1$\kern-.75em\lower1ex%
\hbox{$\sim$}}}}








\def\www#1{\hbox{\tt#1}}

\let\thanks=\footnote
\newcount\fnotenum\fnotenum=0
\def\footnote#1{\advance\fnotenum by 1 \thanks{$^{\the\fnotenum}$}{#1}}

\newcount\itemnum
\let\Item=\item
\def\beginenumerate{\itemnum=0\relax \par\begingroup\nobreak
\def\item{\advance\itemnum by 1 \Item{\the\itemnum.}}}

\def\bitem{\item{$\bullet$}}

\def\titleparagraphs{\interlinepenalty=9999
     \leftskip=0.03\hsize plus 0.22\hsize minus 0.03\hsize
     \rightskip=\leftskip \parfillskip=0pt
     \hyphenpenalty=9000 \exhyphenpenalty=9000
     \tolerance=9999 \pretolerance=9000
     \spaceskip=0.333em \xspaceskip=0.5em }
\def\center#1{\par{
     \def\\{\break} \titleparagraphs \noindent #1\par}}


\def\remark{\global\advance\remarknum by 1
   \smallskip{\it Remark \the\remarknum.}\enspace\ignorespaces}

\font\titlefont = cmbx10 at 12 pt

\def\title#1{\center{\baselineskip=14pt\titlefont\ignorespaces#1}}
\def\author#1{\bigskip\center{\ignorespaces\bf#1}}
\def\address#1{\smallskip\center{\ignorespaces#1}}
\def\date#1{\medskip\centerline{#1}}


\input epsf
\let\label=\autoeqno
\let\ref=\eqref

\def\textbf#1{{\bf#1}}
\def\texttt#1{{\tt#1}}
\def\mbox#1{\hbox{#1}}

\hyphenation{an-a-lyse be-hav-iour}

\def\Ne{N_{\rm e}}
\def\Nreal{N_{\rm real}}
\def\Nbg{N_{\rm bg}}

\def\PB{P_{\rm B}}

\bibitem{we}
   G. V. Kulikov, M. Yu.\ Zotov, \texttt{arXiv:astro-ph/0407138};
   M.~Yu.~Zotov, G.~V.~Kulikov, Izvestiya RAN, ser. fiz., \textbf{68}
   (2004)~1602 (in Russian); {\it ibid.,} \textbf{71} (2007)~502;
   \texttt{arXiv:astro-ph/0610944}.

\bibitem{Al}
   D. E. Alexandreas, D. Berley, S. Biller et al.,
   Astrophys.~J. \textbf{383} (1991) L53.

\bibitem{Green}
   D. A. Green, 2006, `A Catalogue of Galactic Supernova Remnants (2006
   April version),' Astrophysic Group, Cavendish Laboratory, Cambridge,
   United Kingdom\\
   (available at \mbox{\texttt{http://www.mrao.cam.ac.uk/surveys/snrs/}}).

\bibitem{ATNF}
   The ATNF Pulsar Database,
   \mbox{\texttt{http://www.atnf.csiro.au/research/pulsar/psrcat/}};\\
   R.~N.~Manchester, G.~B.~Hobbs, A.~Teoh, M.~Hobbs, Astrophys.~J. \textbf{129}
   (2005)~1993; \texttt{arXiv:astro-ph/0412641}.

\bibitem{Milagro}
	A. A. Abdo, B.~Allen, T.~Aune et al., \texttt{arXiv:0801.3827}

\bibitem{puzzling}
	L. Drury, F. Aharonian, \texttt{arXiv:0802.4403}

\bibitem{Octave}
   J. W. Eaton, ``GNU Octave: A High-Level Interactive 	
   Language for Numerical Computations.'' Edition~3 for version 2.0.13, 1997;
   \mbox{\texttt{http://www.octave.org/}}

\title{Regions of Excessive Flux of PeV Cosmic Rays%
\thanks{$^*$}{\rm Presented at the 30th Russian Cosmic Ray Conference,
 Saint-Petersburg, July 2--7, 2008; \www{http://www.ioffe.ru/rcrc2008/}}}

\author{M. Yu.~Zotov, G. V. Kulikov}
\address{%
   D. V. Skobeltsyn Institute of Nuclear Physics\\
   Moscow State University, Moscow 119992, Russia\\
   \tt $\{$zotov,kulikov$\}$@eas.sinp.msu.ru}

\date{February 10, 2009}

\bigskip
\centerline{\bf Abstract}
{\narrower\noindent
An analysis of arrival directions of extensive air showers generated by
cosmic rays in the PeV energy range and registered with the EAS MSU and
EAS--1000 Prototype arrays reveals a considerable number of regions of
excessive flux of cosmic rays. We present results of comparative analysis
of regions found in the two data sets, estimate probabilities of their
appearance, and discuss correlation of their locations with coordinates
of possible astrophysical sources of PeV cosmic rays.

}


\section{Method of The Investigation}{sec:method}
The problem of the origin of cosmic rays (CRs) in the PeV energy range
remains open for many years now. Attempts to find coincidences between
arrival directions of CRs and their possible astrophysical sources is one
of the routes of investigations on the subject. In the paper, we continue
the earlier investigation~\cite{we} and present some of the results of a
comparative analysis of arrival directions of extensive air showers (EAS)
registered with the EAS MSU array in 1984--1990 and with the EAS--1000
Prototype (``PRO--1000'') array in 1997--1999.

We selected 513~602 showers of the EAS MSU data set and 1~342~340 showers
of the PRO--1000 data set for the purposes of the investigation. Showers
registered with the two arrays have different number of charged
particles~$\Ne$ in a typical event.  For the EAS MSU array, the median
value of~$\Ne$ is of the order of $1.6\times10^5$, while that for the
PRO--1000 array equals $3.7\times10^4$. All selected EAS have zenith
angles $\theta < 45.7^\circ$.

The investigation is based on the method of Alexandreas et~al.~\cite{Al},
which has been used by different research groups for the analysis of
arrival directions of CRs.  The idea of the method is the following.
Every shower in the experimental data set obtains arrival time of another
shower randomly. After this, new equatorial coordinates
$(\alpha,~\delta)$ are calculated for the ``mixed'' data set thus
providing a ``mixed'' map of arrival directions.  The mixed map has the
same distribution in~$\delta$ as the original map.  Both maps are divided
into sufficiently small ``basic'' cells thus providing a way for
comparison.

Mixed maps must be created multiple times in order to minimize the
dependence of the result on the pseudo-random choice of arrival times.
Finally, one calculates the mean of the mixed maps thus obtaining a
``background'' map.  The underlying idea of the method is that this
background map has most of the properties of an isotropic background, and
presents the distribution of arrival directions of cosmic rays that would
be registered with the array in case there is no anisotropy. Thus,
deviations of the real map from the background one may be assigned to a
kind of anisotropy of arrival directions of EAS registered at the array.

In our earlier analysis of arrival directions of EAS registered with
PRO--1000, basic cells had the size $1^\circ\times1^\circ$, and the
number of cycles of ``mixing'' varied from 100 to 200~\cite{we}. In the
present investigation, we reduced the size of basic cells down to
$0.5^\circ\times0.5^\circ$ in order to improve accuracy of locating the
regions of excessive flux (REFs). This led to a slower rate of
convergence of mixing. We thus had to perform a greater number of cycles
of mixing. The difference in the number of EAS in basic cells of two
consecutive maps is $\le0.02$  after 800 cycles for the EAS MSU data set,
and after approximately 900 cycles for the PRO--1000 data set. Thus we
have chosen to run 1000 cycles of mixing for both data sets.

Regions of excessive flux (REFs) of CRs were searched for in the
following way. Adjacent basic strips in~$\delta$ (each $0.5^\circ$ wide)
were joined into strips with width $\Delta\delta=3^\circ\dots30^\circ$
with step equal to~$0.5^\circ$. Each wide strip was then divided into
adjacent cells of width~$\Delta\alpha$. After this, we calculated the
number of EAS inside each of these cells for both experimental ($\Nreal$)
and ``background'' ($\Nbg$) maps. For each pair of cells, we then
calculated the value of significance $S = (\Nreal - \Nbg)/\sqrt{\Nbg}$. A
cell was considered to be a region of excessive flux if $S>3$.

Rather unexpectedly, maps obtained after 1000 cycles of mixing with
different random seeds in different runs lead to selection of slightly
different sets of REFs.  Due to this, we performed three 1000-cycle runs
for the EAS MSU data set and two 1000-cycle runs for the PRO--1000 data
set.  Only those REFs found in all runs were selected for the following
analysis.

The method of Alexandreas et~al.~\cite{Al} does not put any requirements
on the shape or size of regions in experimental and background maps to be
compared. In what follows, we only consider so called ``regular''
regions.  These are regions that have approximately the same area at
different~$\delta$ and  fixed~$\Delta\delta$, and have the shape of a
square at $\delta=0^\circ$. To accomplish this, we used the standard
formula $\Delta\alpha=\Delta\delta/\cos\bar{\delta}$,
where~$\bar{\delta}$ is the mean value of~$\delta$ for the current strip,
and $\Delta\alpha$ is rounded to the nearest half-integer number.

The method of Alexandreas et~al. does not provide a direct answer to the
question about the probability of appearance of a REF. To solve the
problem, we introduced a simple model based on the binomial distribution.
In the model, the number of trials equals the number of showers~$N$ in
the data set under consideration, and an estimation of success (for a
fixed region) equals $\tilde p = \Nbg/N$, where~$\Nbg$ is the number of
showers in the region of the background map. The assumption is based on
the fact that~$\Nbg$ may be considered as an expected number of showers
in the region. Obviously, the probability of finding exactly~$\Nreal$ EAS
in the region equals
\[
	P(\nu=\Nreal) = C(N,\Nreal) \tilde{p}^{\Nreal} (1 - \tilde{p})^{N-\Nreal},
\]
where $\nu$ is a random variable equal to the number of successes in the
binomial model, and $C(N,\Nreal)$ is the corresponding binomial
coefficient. A more interesting probability to be considered is
$\PB=P(\nu\le\Nreal)$.

The condition $S>3$ does not imply that it holds for sub-regions of a
given region. Due to this, we have employed a number of additional
quantities. One of the most useful of them is a probability calculated
from the following binomial model, which characterizes the degree of 
homogeneity of excess of the experimental flux in a given region over the
background one. Let~$\xi$ be a random variable equal to the number of
basic cells of the given region with an excess of EAS (over the
background flux). The number of trials equals the number of basic cells
in the region ($\ge6\times6=36$). It is natural to assume the probability
of ``success'' to be equal to~1/2. A probability to consider is
$P^+=P(\xi<N^+)$, where~$N^+$ is the number of basic cells in the region
with an excess of EAS in the experimental map over the background one.


\section{The Main Results}{sec:results}
Approximately 900 regular regions such that $S>3$, $P^+>0.95$ were found
in the PRO--1000 data set, see Fig.~1.  For all regions,
$\max\Delta\delta=17.5^\circ$, and the overwhelming majority of them have
$\max\Delta\delta\le12^\circ$. Nearly 400 regions that satisfy the above
two conditions were found in the EAS MSU data set, see Fig.~1. For all of
them, $\Delta\delta\le10^\circ$. The probability $\PB\ge0.99812$ for all
REFs in both data sets. The number of EAS inside REFs differs from a few
dozens up to approximately four thousand in the EAS MSU data set and 48.5
thousand in the PRO--1000 data set.

\figure{fig:fig1}{%
	Regular regions of excessive flux such that $S>3$ and $P^+>0.95$
	found in the PRO--1000 (blue) and EAS MSU (red) data sets.
	For the sake of simplicity, joint boundaries instead of all
	individual REFs are shown in a number of cases.
   The arcs at the left and the right sides of the figure
	show the Galactic plane. The $\cap$-like curve shows the Supergalactic plane.
}{
	\epsfxsize=0.61\hsize \epsfbox{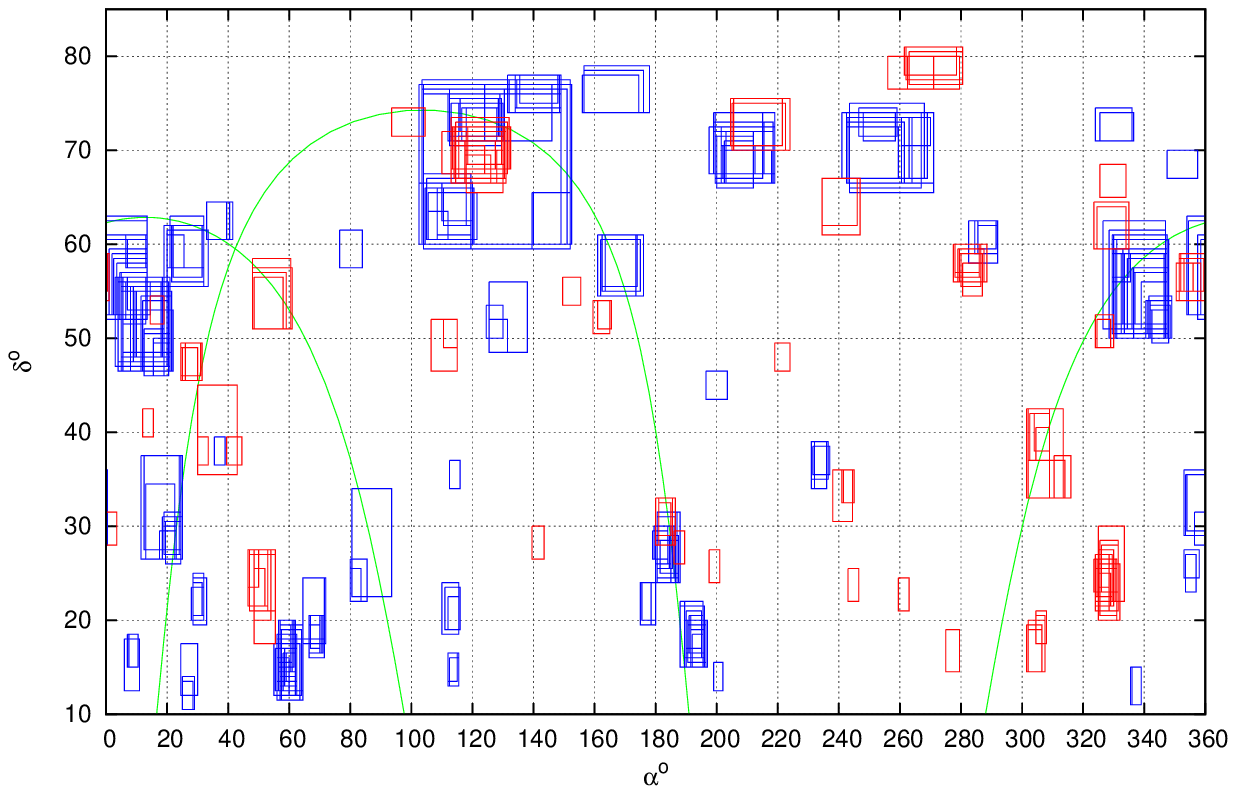}
}

It is clearly seen that REFs found in the EAS MSU data set are, as a
rule, more compact than that in the PRO--1000 data set.  This may be
related to the better accuracy in determination of arrival directions
with the EAS MSU array. It is also clear that the number of coincidences
or ``embeddings'' of REFs found in two data sets is comparatively small.
In our opinion, the most interesting of them are the following:

\bitem a number of intersecting regions near the Supergalactic plane around 
$\alpha=120^\circ, \delta=70^\circ$ and $\alpha=183^\circ, \delta=30^\circ$;

\bitem a number of intersections of REFs located near the Galactic plane
at $\alpha\sim323^\circ$--$335^\circ$, and $\alpha\sim350^\circ$--$360^\circ$;

\bitem an embedding of a REF from the EAS MSU data set in a huge REF of
the PRO--1000 set in the vicinity of $\alpha=17^\circ, \delta=53^\circ$;

\bitem the only embedding of a REF from the PRO--1000 data set in a much
larger EAS MSU region in the vicinity of $\alpha=37^\circ, \delta=38^\circ$.

Intersections of REFs located around $\alpha=215^\circ, \delta=73^\circ$
and $\alpha=285^\circ, \delta=58^\circ$ far from both Galactic and
Supergalactic plans seem to be remarkable, too.  One can also point out a
number of pairs of REFs located close to each other.  On the other hand,
many REFs do not have ``neighbours'' in the other data set. In
particular, there are no regular REFs satisfying the above criteria in
the PRO--1000 set in the vicinity of the Galactic plane in the direction
close to the Galactic center (the lower part of the arc at the right-hand
side of Fig.~1).

Let us consider a relation between location of the REFs and coordinates
of two traditional groups of possible Galactic sources of cosmic rays
in the PeV energy range, namely supernova remnants (SNRs) and pulsars,
see Fig.~2.

\figure{fig:fig2}{%
	Regular REFs found in the EAS MSU (upper figure) and PRO--1000 data sets
	and coordinates of Galactic SNRs~($\bullet$)~\cite{Green} and pulsars
	($\circ$ for pulsars with $d\le3$~kpc,
	$\triangle$ for those with $3<d\le6$~kpc,
	$\times$ for the other)~\cite{ATNF}.
}{
	\epsfxsize=0.61\hsize \epsfbox{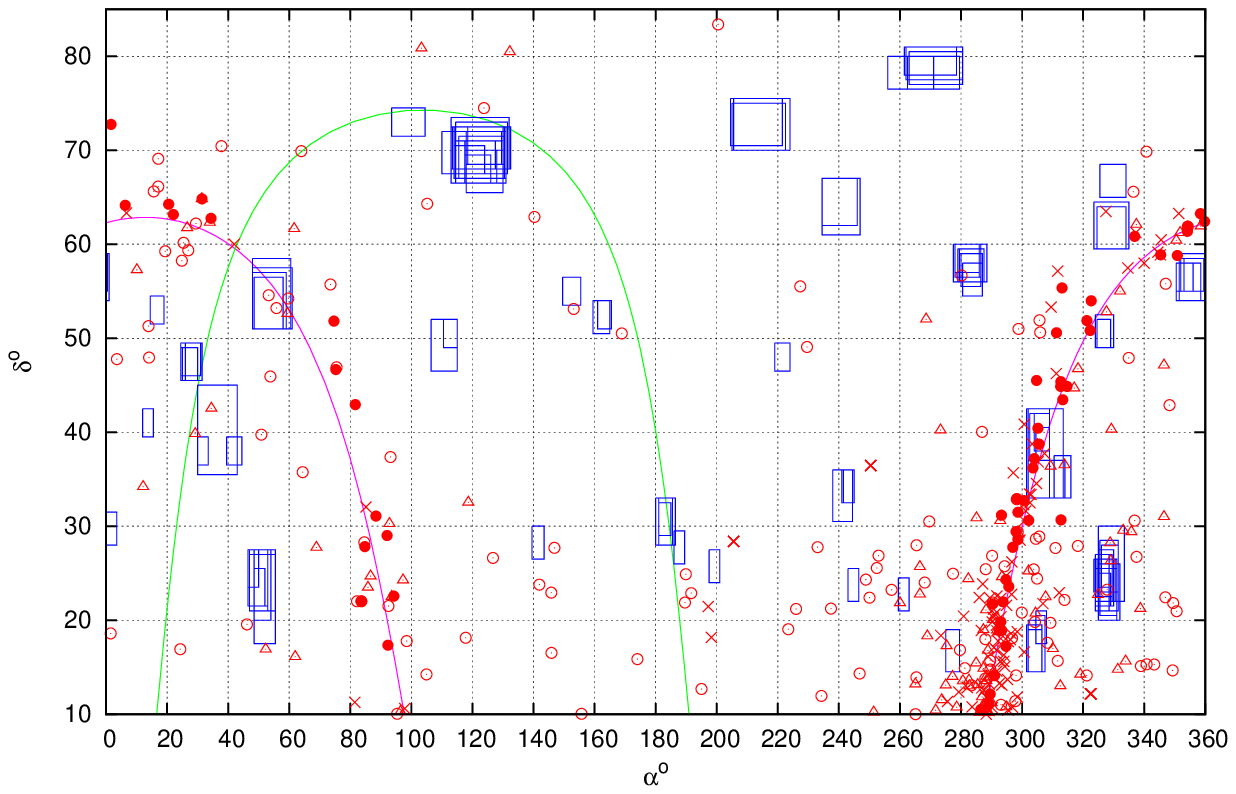}
	\epsfxsize=0.61\hsize \epsfbox{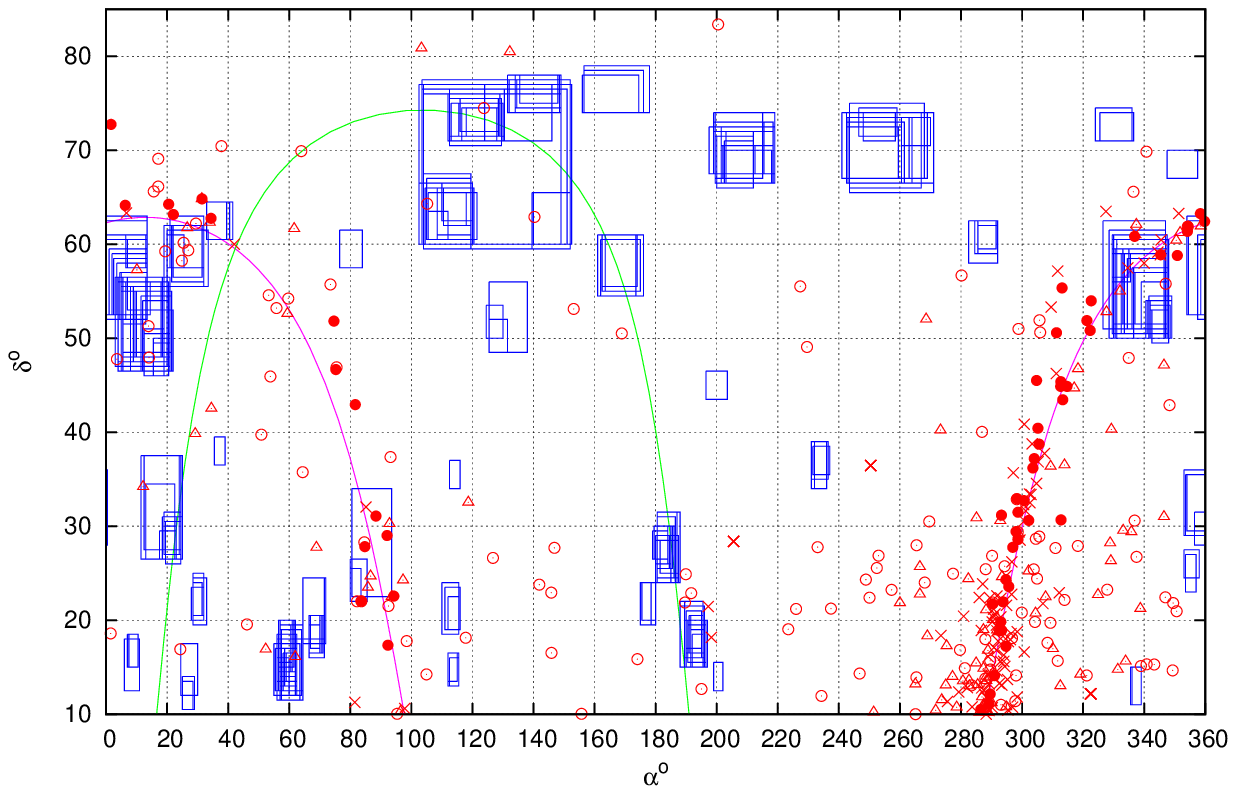}
}

Coordinates of the following nine SNRs are located inside or in the near
vicinity of REFs from the EAS MSU data set (from top to bottom along the
right-hand arc of the Galactic plane): G106.3+2.7, Cassiopeia~A (3C461),
CTB~104A, 3C434.1, $\gamma$-Cygni, G76.9+1.0, CTB~87, G73.9+0.9 and
G69.7+1.0. Almost all of them are of the shell type. For a considerable
part of REFs in the EAS MSU set, there are pulsars with close coordinates
and located at distances $d=0.33\dots3$~kpc from the Solar system.  Among
them, one can specify
J1840+5640 ($d=1.70$~kpc),
J1838+1650 ($d=1.84$~kpc),
J1709+2313 ($d=1.83$~kpc),
J1012+5307 ($d=0.52$~kpc),
J1115+5030 ($d=0.54$~kpc),
J0814+7429 ($d=0.33$--0.43~kpc),
J0332+5434 ($d=1.06$--1.44~kpc) and a few other.
At the same time, a few groups of REFs in the EAS MSU data set (including
some of those with intersections with REFs in the PRO--1000 set) have
neither SNRs, nor pulsars in their vicinity.

As we have already reported earlier~\cite{we}, the situation for the REFs
in the PRO--1000 data set is qualitatively the same. Namely, there are a
whole number of Galactic SNRs  (e.g., Cassiopeia~A, the Crab Nebula
(SN~1054), S147, 3C157, Tycho's Supernova (SN~1572), SN~1181 and
G132.7+1.3) and pulsars in the vicinity of REFs but still a number of
regions do not have any of the above two classes of sources nearby. One
of such REFs is a region located within
$\alpha=64.5^\circ\dots72.0^\circ$, $\delta=16^\circ\dots24.5^\circ$. It
is especially remarkable because its boundaries are pretty close to the
boundaries of an unidentified REF of 10-TeV cosmic rays  found with the
Milagro array~\cite{Milagro}. To the best of our knowledge, the origin of
this REF has not been explained  yet~\cite{puzzling}.

\section{Conclusions}{sec:conclusions}
An analysis of arrival directions of EASs generated by cosmic rays in the
PeV energy range and registered with the EAS MSU and EAS--1000 Prototype
arrays reveals a considerable number of regions of excessive flux. Many
of them have Galactic supernova remnants and pulsars located nearby but a
number of REFs have none of the two types of cosmic ray sources in their
vicinity.

\medskip
Only free, open source software was used for the investigation.
In particular, all calculations were performed with
GNU Octave~\cite{Octave} running in Slackware Linux.
The investigation was partially supported by Federal contract
No.~02.518.11.7073 and the Russian
Foundation for Fundamental Research grant No.~08-02-00540.
\listrefs
\bye